\documentclass[nonacm, screen, acmlarge]{acmart}

\usepackage{todonotes}
\AtBeginDocument{%
  }

\setcopyright{acmlicensed}
\copyrightyear{2018}
\acmYear{2018}
\acmDOI{XXXXXXX.XXXXXXX}

\acmConference[Conference acronym 'XX]{Make sure to enter the correct
  conference title from your rights confirmation emai}{June 03--05,
  2018}{Woodstock, NY}
\acmISBN{978-1-4503-XXXX-X/18/06}

\begin{document}

\title{EarCapAuth: Biometric Method for Earables Using Capacitive Sensing Eartips}


\author{Richard Hanser}
\affiliation{%
  \institution{Karlsruhe Institute of Technology}
  \city{Karlsruhe}
  \country{Germany}}
\email{hanser@teco.edu}

\author{Tobias R\"{o}ddiger}
\affiliation{%
  \institution{Karlsruhe Institute of Technology}
  \city{Karlsruhe}
  \country{Germany}}
\email{tobias.roeddiger@kit.edu}

\author{Till Riedel}
\affiliation{%
  \institution{Karlsruhe Institute of Technology}
  \city{Karlsruhe}
  \country{Germany}}
\email{till.riedel@kit.edu}

\author{Michael Beigl}
\affiliation{%
  \institution{Karlsruhe Institute of Technology}
  \city{Karlsruhe}
  \country{Germany}}
\email{michael.beigl@kit.edu}

\renewcommand{\shortauthors}{Richard Hanser, Tobias R\"{o}ddiger, Till Riedel, \& Michael Beigl}

\begin{abstract}
Earphones can give access to sensitive information via voice assistants which demands security methods that prevent unauthorized use. 
Therefore, we developed EarCapAuth, an authentication mechanism using 48 capacitive electrodes embedded into the soft silicone eartips of two earables. 
For evaluation, we gathered capactive ear canal measurements from 20 participants in 20 wearing sessions (12 at rest, 8 while walking).
A per user classifier trained for authentication achieves an EER of 7.62\% and can be tuned to a FAR (False Acceptance Rate) of 1\% at FRR (False Rejection Rate) of 16.14\%. 
For identification, EarCapAuth achieves 89.95\%.
This outperforms some earable biometric principles from related work.
Performance under motion slightly decreased to 9.76\% EER for authentication and 86.40\% accuracy for identification.
Enrollment can be performed rapidly with multiple short earpiece insertions and a biometric decision is made every 0.33s.
In the future, EarCapAuth could be integrated into high-resolution brain sensing electrode tips.

\end{abstract}

\ccsdesc[500]{Human-centered computing~Empirical studies in ubiquitous and mobile computing}
\ccsdesc[500]{Security and privacy~Biometrics}

\keywords{earables, hearables, capacitive sensing, authentication, earphones}
\begin{teaserfigure}
  \includegraphics[width=\linewidth,page=2,trim=0cm 11.5cm 0cm 0cm]{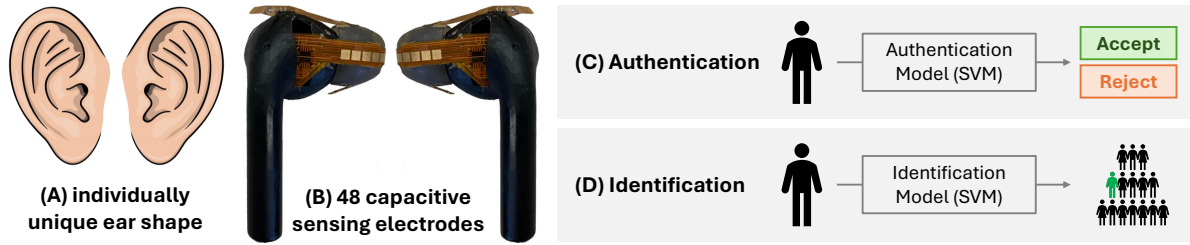}
  \caption{(A) Humans can be identified by unique anatomic properties of the ear \cite{purkait2016external, orkk}; (B) custom earpieces (left, right) with 48 capactive electrodes (2 $\rm\times$ 24) measure the shape of the ear canal of the user; (C) based on the capacitive sensing signals, a classifier accepts or rejects an enrolled user; (D) another classifier can identify a user from a database of enrolled users.}
  \Description{This figure depicts the concept of the ear-based biometric authentication and identification system. Panel A shows two illustrations of human ears, captioned as "individually unique ear shape," emphasizing that each person's ear shape is distinct and can be used as a biometric marker. Panel B shows an image of a pair of in-ear devices with "48 capacitive sensing electrodes" embedded within them. These electrodes measure the ear’s physical features via capacitive sensing for for biometric purposes. Panel C illustrates the authentication process. The system uses a support vector machine (SVM) as the "Authentication Model." The process includes input from a person, which is passed through the model. The system either "Accepts" or "Rejects" the input based on whether it matches a pre-registered user. Panel D illustrates a process for identification. Similar to Panel C, an individual’s input is passed through an "Identification Model" (also using SVM). However, in this case, the model identifies the person from a larger group of people, as illustrated by a group icon. One individual is highlighted as being identified.}
  \label{fig:teaser}
\end{teaserfigure}


\maketitle

\section{Introduction}
\label{sec:Introduction}

Headphones and earphones are becoming increasingly powerful and commercial devices already offer access to sensitive information about the wearer, for example, via voice assistants \cite{apple_siri_airpods_2024}.
In addition, the so-called ``earables'' are a new class of devices which come equipped with sensing capabilities to collect and store sensitive information about the wearer \cite{revi}, for example, longitudinal health \cite{roddiger2021detecting} or movement activity \cite{ferlini2021eargate}. In addition, mixing input from multiple users or by malicious actors will endanger the integrity of the data and lead to erroneous action. 

Most computing devices today are secured by passwords or PINs.
Although methods for entering pins on earphones have been reported in previous research \cite{budsauth, smartear}, they can be time-consuming for the user as they have to be re-entered each time they interact with a device. 
In addition, the input can potentially be observed by attackers. 
A possible alternative could be biometric authentication systems that have been increasingly used in recent years in smartphones such as Apple's FaceID or TouchID \cite{bima}.
However, for earables, the established biometric principles used with smartphones are not suitable as the devices neither have a clear view of the user's face nor are typically used with the finger. 
As another alternative voice recognition may be used. However, it is difficult to use in loud environments, requires significant computing power and enrollment time, and is vulnerable to replay attacks \cite{replay}. 

Therefore, as an alternative approach, we propose EarCapAuth, which leverages the unique shape of the user's ear, which is a biometric feature that is individually different \cite{eov2}.
We implement a custom device based on OpenEarable \cite{oe} that integrates 48 capacitive electrodes into the soft silicone tip of two earpieces (24 left and right each). 
We open-source the hardware under an MIT license.
In a user study, we gather data from 20 participants in 20 wearing sessions.
Based on the dataset, we develop a machine learning pipeline that can authenticate the wearer by making an accept or reject decision. 

In this paper, we present an open hardware prototype and a machine learning pipeline that could also be integrated with high-resolution in-ear EEG electrodes, e.g., as recently patented by Apple \cite{azemi2023biosignal}. 
Our prototype already achieves an equal error rate (ERR) of 7.62\%. At a false acceptance rate (FAR) of 1\%, EarCapAuth achieves a false rejection rate (FRR) of 16.14\%.
We implement another pipeline for identification in which the target user should be identified from the dataset and achieve 89.95\% accuracy.
EarCapthAuth is designed to perform reasonably well under motion-induced stress achieving an EER of 9.76\% for authentication and an accuracy of 86.40\% for identification.
EarCapAuth makes a biometric decision every 0.33 seconds and enrollment can be performed with a few short insertions of the earpiece into the ear canal.

In sum our contributions are: (i) EarCapAuth: an authentication principle for earables based on 48 sensing electrodes embedded into the soft silicone tip inside the ear canal (24 per ear) and (ii) an evaluation of the approach with 20 participants in 20 wearing sessions each showing that our approach works reliably, even under motion-induced stress; (iii) an open-source, MIT-licensed capacitive sensing eartip hardware compatible with OpenEarable \cite{oe}.

\section{Background and Related Work}
\label{sec:background-and-related-work}

In the following, we introduce the theoretical background and related work of ear biometrics, earable authentication and identification, and capacitive sensing,including its application for authentication methods. 

\subsection{Earable Biometrics}
The human ear begins to develop between the fifth and seventh week of
pregnancy. In the fourth month after birth, the shape is fully developed, followed only by growth in size \cite{johnson2019growth}. 
\citet{purkait2016external} analyzed the ears of 2661 individuals and found that 99.9\% of the ears occupy a distinct position in a multi-dimensional feature space. Taking both ears of a person into account left none of the individuals undistinguished from the others.
Therefore, using the shape of the ear for authentication and identification purposes is a viabale and fundamental problem of pattern recognition, which has been explored in many computer vision publications \cite{abaza2013survey}. 
For vision-based ear recognition, there are benchmark datasets \cite{hoang2019earvn1}. However, vision-based approaches are impractical for the user in earphones as they rarely can get a view of the ears and a camera requires significant processing power and energy, which is unlikely to be available state-of-the-art battery-operated earphones (29 - 155 mAh) \cite{varta_lithium_ion_button_cells}. 
Hence, an alternative principle for authenticating users when using earphones is desirable.

Since individual features of the shape of the ears provide an opportunity for biometric authentication \cite{orkk}, several methods for earable authentication using ear biometrics have been proposed. 
To circumvent a camera, related works commonly measure acoustic phenomena inside the ear canal, such as the individual reflection of test signals inside the ear canal (e.g., EarEcho \cite{earecho}, EarDynamic \cite{eardynamic}, MetaEar \cite{metaear}), the individual sound transmission from the throat to the ear canal (e.g., Voice In Ear \cite{voiceinear}), the individual sound produced by teeth occlusions (e.g., Toothsonic \cite{toothsonic} and TeethPass \cite{teethpass}), the unique acoustic pattern of fingers rubbing on the face (e.g., EarSlide \cite{earslide}), or footsteps propagated into the ear canal (e.g., EarGate \cite{ferlini2021eargate}).
Other alternatives based on non-acoustic features include individual in-ear EEG patterns \cite{eeg} or unique changes in in-ear PPG measurements during speaking \cite{ppg}. 
However, to the best of our knowlege, none have used capacitive sensing embedded into earphones to assess the shape of the ear canal and auricle. 

\subsection{Capacitive Sensing and Biometrics}
Capacitive sensors are pervasive in human-computer interaction as they give good results with simple technology \cite{commgr}. 
Capacitive sensing is based on the principle that two electrodes with a voltage difference always have a capacitance between them, which is dependent on the electrodes, the distance and the environment between them \cite{commgr}. By measuring this capacitance, it is possible to indirectly measure the (variable) distance between two electrodes under otherwise constant conditions. Capacitance is usually measured by applying a defined current for a defined time span and measuring the voltage reached \cite{commgr}. By varying the current or time it is possible to cover a wide measurement range \cite{mpr}.  

A widespread application of capacitive sensing in authentication are fingerprint sensors, which contain a high-resolution array of electrodes to detect the small differences in height of the fingertip \cite{fprint}.
Another example of capacitive sensing biometrics is BodyPrint technology \cite{bodypr}. It uses the imprint of body parts, including the ears, measured via a capacitive touchscreen to identify users. In contrast, our work explores earable authentication using capacitive electrodes embedded into the tip of earphones.

Research on earables that has explored capacitive sensing includes iSkin \cite{weigel2015iskin} for interaction using a capacitive touch surface on the backside of the auricle. Similarly, \citet{lissermann2013earput} introduced EarPut, a series of capacitive surfaces for ear-based touch and mid-air interaction. In addition, capacitive detection can be used to detect if the user is currently wearing an earpiece \cite{buil2005headphones}.
To the best of our knowledge, no previous work has explored capacitive electrodes embedded in the ear canal for biometric security applications.

\section{EarCapAuth}
\label{sec:earcapauth}

\begin{figure}[!t]
    \centering
    \includegraphics[width=\linewidth,page=1,trim=0cm 10.5cm 0cm 0cm]{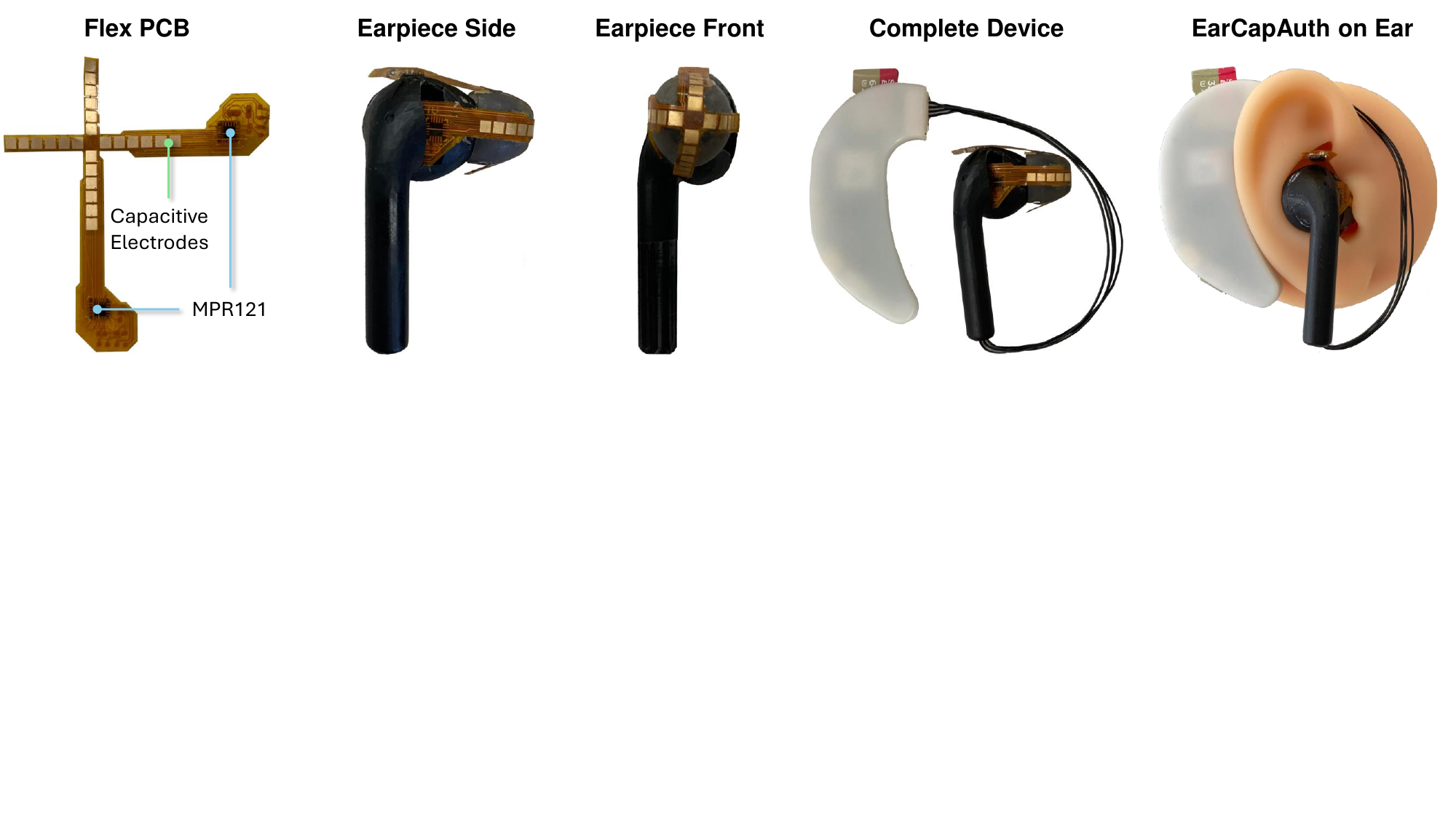}
    \caption{The custom-built device used to realise the EarCapAuth authentication method. Each earbud (left or right) contains the two MPR121 capacitive sensor connected to 8 electrodes each. The flex printed-circuit board (flex PCB) is folded and glued onto a soft silicone eartip. A 3D-printed enclosure houses the components. The earpiece is connected via four cables to the JST connector of OpenEarable \cite{oe} which records capacitive sensing data to an on-board microSD card.}
    \Description{The figure illustrates the components and assembly of the custom-built device used for "EarCapAuth." It progresses through different stages, starting with the Flex PCB on the far left, which shows a flexible printed circuit board (PCB) with capacitive electrodes and an MPR121 chip. These capacitive electrodes form part of the biometric sensing system, while the MPR121 functions as a controller for capacitive sensing. The capacitive sensing electrodes are arranged in a cross shape with 6 electrodes on each arm. The arms fold onto the silicone tip of an earpiece. The Earpiece Side view, shows the flexible PCB and capacitive electrodes integrated into the device. This is followed by the Earpiece Front view, where the earpiece is displayed from the front, giving a clearer look at how the capacitive sensing electrodes are arranged to interact with the user's ear the electrodes spraed from the tip at the center into the four cardinal directions towards the back of the silicone tip. The Complete Device view follows, displaying the full assembly of the ear-based device. This image includes a curved, white external component, which is OpenEarable, the device used to read out the capacitive electrodes. Finally, the EarCapAuth on Ear view demonstrates the device fitted onto a mock ear, showing how the complete system looks when worn.}
    \label{fig:hardware}
\end{figure}

We introduce EarCapAuth, an authentication technique based on capacitive shape sensing of the ear canal surface. A custom earpiece was implemented based on OpenEarable \cite{oe} and commercial of-the-shelf components with 2 $\rm\times$ 24 electrodes placed in the left and right ear canal, respectively, see \autoref{fig:teaser} and \autoref{fig:hardware}. We make our custom earpiece hardware available as an MIT-licensed open-source design. A feature-based machine learning classifier was implemented to perform authentication and identification, see \autoref{fig:classifier}.

\subsection{Concept}
As the shape of the ear canal is a unique biometric feature \cite{orkk, purkait2016external}, we employ capacitive sensing to measure its shape. 
Individual bending and distances between the 24 electrodes per ear result in measurements that correlate with the anatomical features of a user.
Choosing a high resolution of electrodes produces a detailed mapping of the ear canal. 
By training a classifier on data of a target user vs. non-target users or a classifier to predict individual users, EarCapAuth can be used for authentication and identification. 

\subsection{Hardware and Firmware} The EarCapAuth device is divided into two parts: a main unit and a custom-made earpiece. 

The foundation and main unit of our sensing device is the OpenEarable \cite{oe} Version 1.3, an extensible platform for earphone hardware, which we use as the main unit. The OpenEarable main unit contains the microcontroller (nRF52840 with ARM Cortex M4), peripherals, additional sensors (not used in our setting), BLE communication and a rechargeable battery. The main OpenEarable unit is worn behind the ear. To realize EarCapAuth, we have implemented a custom earpiece for OpenEarable (\autoref{fig:hardware}), which is connected to the main unit via an 8-pin JST connector. The earpiece consists of a custom-designed flexible PCB that is mounted on a regular silicone earpiece tip using power glue. 
A total of 24 electrodes for capacitive measurements are printed on the printed circuit board (PCB). These two PCBs are distributed across four arms (six electrodes each), forming a cross-shaped configuration. This cross is then adhered to the speaker earpiece. The electrodes on each PCB are connected to an MPR121 capacitive sensor controller, which includes internal filtering \cite{mpr}. This controller, together with the electrodes, is mounted on the Flex PCB to minimize ambient noise induction and increase signal stability. The two sensor controllers, one for each PCB, are individually connected to the OpenEarable via I²C over the connection cable. The internal battery of the OpenEarable was used to power the OpenEarable and the earpiece to avoid any connection to ground or other objects. 
We built one for the left ear and one for the right ear, bringing the total number of electrodes to 48. 
All electrodes are sampled at 15 Hz.
To sync the left with the right ear device, the OpenEarable button is pressed three times at the beginning and end of the study.

We extend the firmware of OpenEarable which is written in C with a capacitive sensor driver. The application software is configured to collect the readings from the capacitive sensors and to store them in a CSV file on a microSD card inserted into OpenEarable.

\subsection{Machine Learning Classifier}
To perform the authentication and identification task with EarCapAuth, two simple linear support vector machine (SVM) models were implemented. 
Such lightweight models could easily be implemented directly on state-of-the-art earphones for edge computation. 
Both models were implemented using Python and scikit-learn \cite{scikit} for training and testing. The evaluation software was run on a standard x64 PC with an Intel i7-13700K and 32GB RAM (DDR4-3200).

\subsubsection{Authentication} For authentication, a linear SVM model is trained on data of the target user labeled as ``accept`` and of non-target users as ``reject''. 
As input, the model takes the arithmetic mean per electrode of 5 sample frames of the left and right ear (= 1 chunk). 
This is equivalent to 0.33 seconds.
The model predicts a probability to accept. 
A configurable threshold parameter can be fine tuned to increase or decrease the false acceptance rate (FAR) and false rejection rate (FRR) of EarCapAuth.
Through hyperparameter grid search, the regularization factor C of the SVM was set to 0.025.

\subsubsection{Identification} Similarly to the authentication task, the arithmetic mean per electrode is computed over the arithmetic mean of five sample frames (= 1 chunk). 
This is equivalent to 0.33 seconds.
For identification, the model is trained on all users that should be in the identification database. 
The model is trained on data of all users labeled as ``User 1`` to ``User N``, whereas N denotes the number of users in the identification database. 
The model predicts the specific user.
Through hyperparameter grid search, the regularization factor C of the SVM was set to 0.025.

\begin{figure}[!t]
    \centering
    \includegraphics[width=\linewidth,page=3,trim=0cm 10.5cm 0cm 0cm]{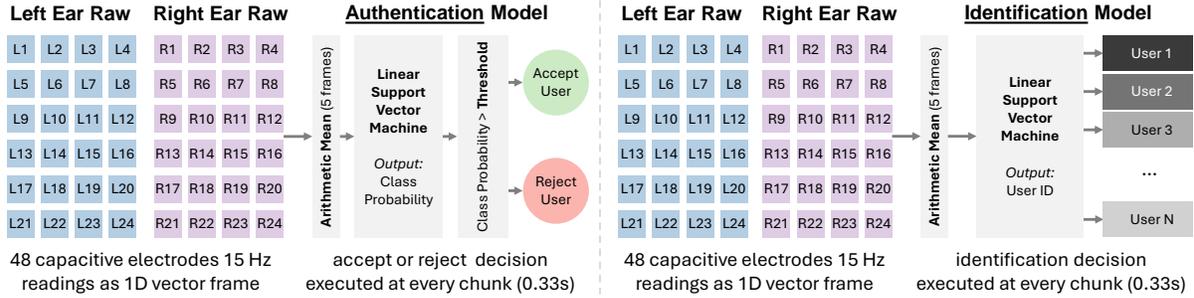}
    \caption{The left ear (L1-L24) and right ear (R1-R24), capacitive signals are used to generate a raw data frames as 1D vectors at 15 Hz. For authentication a linear support vector machine (SVM) makes an accept or reject decision for user authentication based on the arithmetic mean of five samples frames. The identification model also uses a linear SVM as classifier. It predicts the respective user from the dataset (User 1, User 2, User 3, etc.).}
    \Description{The figure illustrates the process of both authentication and identification using ear biometric data captured from capacitive electrodes. On the left side, the raw data from 48 capacitive electrodes, 24 from each ear (left and right), are collected at a rate of 15 Hz. These readings are represented as a 1D vector frame with individual sensor data points labeled as L1-L24 for the left ear and R1-R24 for the right ear. In the authentication process, the arithmetic mean of 5 frames is calculated, and this information is fed into a linear Support Vector Machine (SVM) model, which outputs a class probability. If the class probability exceeds a predefined threshold, the user is authenticated and accepted; otherwise, the user is rejected. The decision to accept or reject is made every 5 frames (0.33 seconds). Similarly, for identification, raw data from the left and right ear electrodes are processed in the same manner, and the arithmetic mean of 5 frames is sent to a linear SVM model, which outputs the predicted user ID. This process allows for identification among multiple users, and the identification decision is also executed every 5 frames (0.33 seconds).}
    \label{fig:classifier}
    
\end{figure}
\section{Evaluation}
\label{sec:evaluation}
In the following we describe how we evaluated EarCapAuth.
The primary objective was to determine the efficacy of the system for authentication and identification in both static \textit{and} dynamic conditions, where signal noise is introduced by user and device movement. 

\subsection{Participants and Apparatus}
In total, 20 participants (16 male, 4 female, between 18 and 50 years) were recruited for the study through a sample of convenience.
As an apparatus, we apply two EarCapAuth devices, one worn on each ear.
The device is carefully cleaned and disinfected after every participant using a cotton swab and isopropyl alcohol. To avoid an allergic reaction, we waited until all the isopropyl alcohol had diffused before using it again.

\subsection{Study Design}
Participation in the study was voluntary and all participants provided their informed consent prior to participation. 
The study adhered to the guidelines formulated in the Declaration of Helsinki and complied with all university ethical guidelines. 
Ethics approval was not specifically required for this type of study within the jurisdiction where the experiment was conducted, in accordance with the university's policies.

The participants wore the EarCapAuth devices in both ears. 
Participants underwent 20 wearing sessions in which the data were captured for 60 seconds per wearing session. 
Between each session, participants take a break, remove the device from their ears, and move.
This should ensure that our experiments can show that EarCapAuth works reliably for different wearing sessions in which the device may be positioned differently.
In addition, our goal was to avoid robotic-like repetition of movements when sitting.
During the first six sessions, we asked participants to remain seated, while they were allowed to use their computer or smartphone.
To test whether EarCapAuth remains reliable under motion-induced signal noise conditions, participants were then asked to stand up and walk around the room during the next four sessions. 
The same process was applied to the next 10 wearing sessions.
During the study, speaking was not allowed. 
As signal quality improves when a person is grounded, it was ensured that study participants were not electrically grounded to avoid unrealistic measurement conditions. The participants were rewarded with small sweet treats.

\subsection{Data Preparation}
In total twenty one minute capacitive electrode recordings were collected at 15 Hz for each participant. 
As the capacitive readings take some time to settle in, we cut off the first 15 seconds, and the last 5 seconds to avoid interference caused by engaging to remove the device.
As described in the pipeline, we split each wearing session into chunks of 5 sample frames each without overlap (40 / 5 = 8~chunks per wearing session).
We apply the arithmetic mean to the 5 $\rm\times$ 48 electrode readings per chunk to smooth out random noise, which is then passed to the classifier (see \autoref{fig:classifier}).
This leaves a total of 20 participants $\rm\times$ 20 wearing sessions $\rm\times$ 8 chunks = 3,200 chunks for evaluation.

\begin{figure}[!t]
    \centering
    \includegraphics[width=\linewidth]{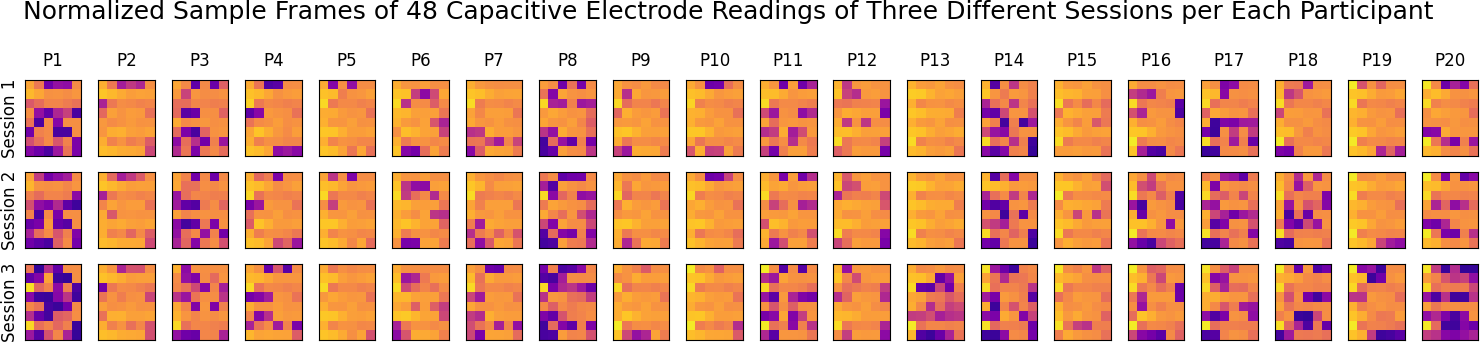}
    \caption{The heatmaps each show the mean of 48 capacitive electrode readings of one chunk (arithmetic mean of 5 samples per channel). They were selected randomly from three different wearing sessions for every participant. 
    The heatmaps show the consistency within participants and the variation between different participants. The color scale represents the electrode activity, with values normalized across all participants and sessions shown in the figure.}
    \Description{This figure presents a series of heatmaps displaying the average electrode activity from 48 capacitive sensors for 20 participants (P1 to P20). The heatmaps are organized in a grid, where each participant has three heatmaps corresponding to three different wearing sessions. These sessions are labeled along the left as "Session 1," "Session 2," and "Session 3." Each heatmap is a small square grid composed of shades of purple and orange, where the color scale represents normalized electrode activity, with purple indicating lower activity and orange representing higher activity. The figure illustrates the consistency of electrode readings within individual participants across the three sessions, as well as the variation between different participants.}
    \label{fig:sample-frames}
\end{figure}

\autoref{fig:sample-frames} are examples chunks of different wearing sessions recorded from all 20 participants. It can be clearly observed that there is a strong coherence across the chunks of individual participants but also significant difference between chunks of different participants. This suggests that the capacitive electrodes are a good foundation for a biometric principle.

\subsection{Training}
In the following, we introduce the training setup for the authentication and identification task for the two pipelines introduced in \autoref{sec:earcapauth}.

\subsubsection{Authentication} For the authentication task, we apply stratified leave-4-sessions-out validation. 
Initially, we only used the chunks from the 12 wearing sessions, when the users were at rest.
We use 8 wearing sessions of the target user for training labeled as ``Accept'' and 8 wearing sessions each of the remaining 19 non-target users labeled as ``Reject''. 
The remaining sessions are used for testing (4 sessions per user).
Consequently, none of the sessions used during training are part of the test set which replicates a real world use case in which the users insert the device freshly.
The model then predicts a class probability to ``Accept''.
A model based on this evaluation strategy is trained for each of the 20 users.
We vary the accept threshold to test how the false acceptane rate (FAR) and false rejection rate (FRR) can be a trade-off against each other depending on the desired properties of the authentication model.

\subsubsection{Identification}
For identification, we apply stratified leave-one-session-out cross-validation. Initially, we only used the data from when users were at rest (12 wearing sessions). To evaluate the identification model, we use 11 wearing sessions of each user for training and 1 wearing session from each user for testing. We perform this step for each of the 12 folds resulting in stratified leave-one-session-out cross-validation. None of the sessions used during training are part of the test set which replicates users inserting the earphones freshly every time they want to be identified.

\section{Results}
\label{sec:results-and-discussion}

We describe the results of the authentication and identification task. We further discuss the impact of the number of wearing sessions used for enrollment and the impact of movement artifacts on the performance of EarCapAuth.

\subsection{Authentication}
In this section, we summarize the authentication performance and the impact of motion artifacts on the authentication performance of EarCapAuth.

\subsubsection{Overall Performance} For authentication, in our study EarCapAuth achieved an equal error rate of 7.62\%. 
The threshold can be configured to fine-tune the authentication model to different FAR and FRR. For example, at 1\% false acceptance rate, the false rejection rate of EarCapAuth is 16.14\%. For comparison, the NIST study of professional 3-inch biometric fingerprint authentication systems, such as those used for border control, showed similar performance (4 out of 5 with 1\% false acceptance) \cite{watson2009slapssegii}, while smaller 2-inch systems had lower performance (3 out of 5). The relationship of FAR vs. FRR for authentication with EarCapAuth is shown in \autoref{fig:results} (A).

\subsubsection{Impact of Movement} When the user moves, the authentication performance is likely to be degraded.
Therefore, we trained the authentication model on the wearing sessions when the users were at rest. 
We then evaluated the authentication model on the data when the participants were moving. 
This only slightly reduced the equal error rate to 9.76\%.
This suggests that EarCapAuth is relativley stable against movement artifacts.

\subsection{Identification}
We present identification performance and provide an analysis of how the amount of training data used during enrollment influences results, as well as the impact of users moving during authentication on performance degradation.

\subsubsection{Overall Performance} 
\begin{figure}
    \centering
    \includegraphics[width=\linewidth,page=4,trim=0cm 8cm 0cm 0cm]{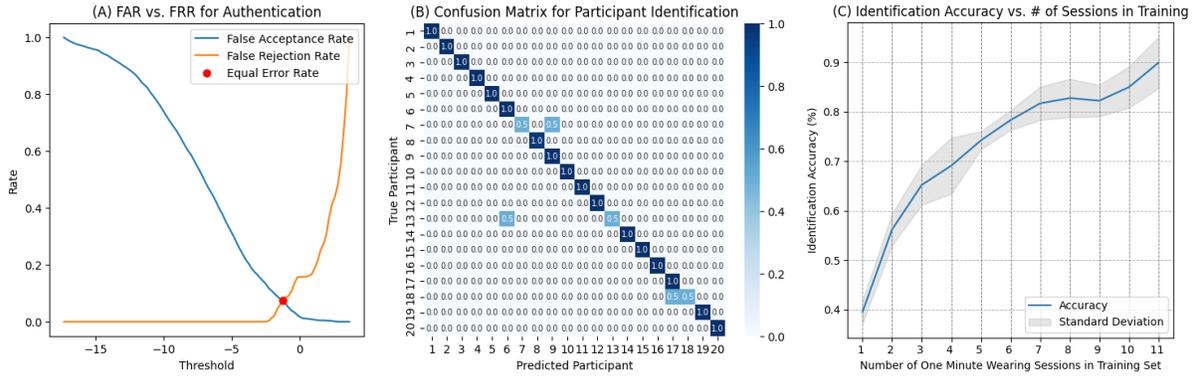}
    \caption{(A) Changing the acceptance threshold value allows fine tuning the model to allow higher false acceptance or false rejection rates; (B) Confusion matrix of one fold of the identification task for 20 participants shows high accuracy for most participants with consistent confusions between participants and wearing sessions; (C) Identification accuracy improves with more training.}
    \Description{This figure presents three different visualizations that collectively illustrate aspects of participant identification and authentication performance. In subplot (A), the graph displays the relationship between the False Acceptance Rate (FAR) and the False Rejection Rate (FRR) as a function of the threshold value used for authentication. As the threshold increases, the FAR (blue line) decreases, while the FRR (orange line) increases, highlighting the trade-off between these two metrics. The red dot marks the Equal Error Rate (EER), the point where the FAR and FRR intersect, indicating the threshold at which both error rates are equal. In subplot (B), a confusion matrix shows the performance of the identification model across 20 participants. The matrix predominantly features high values along the diagonal, signifying that the model accurately identifies most participants. However, some off-diagonal elements suggest occasional misclassifications, where the model confuses participants. Subplot (C) illustrates the relationship between identification accuracy and the number of one-minute training sessions. As the number of training sessions increases, the model's identification accuracy improves, with the accuracy reaching approximately 90\% after 11 sessions. The shaded area in this plot represents the standard deviation, indicating variations in performance. Overall, the figure demonstrates how adjusting the threshold impacts authentication error rates, the accuracy of participant identification, and how additional training data leads to improved identification performance.}
    \label{fig:results}
\end{figure}

EarCapAuth achieved an overall identification accuracy of 89.95\% $\pm$ 5.17\%. The confusion matrix of one fold is shown in \autoref{fig:results} (B). Upon further investigation, we find that the miss-classifications are relativley consistent between the same individuals and wearing sessions across folds. This suggests that some participants have similar ear shapes which are harder to distinguish for the model.

\subsubsection{Enrollment Duration} 
Similar to how commercial fingerprint or face scanning methods request multiple scans for enrollment, we noticed that with an increasing number of wearing sessions used during training, the performance of EarCapAuth improves.
To systematically understand this effect, we evaluated the performance of EarCapAuth identification by incrementally increasing the number of wearing sessions.
The results are summarized in \autoref{fig:results} (C).
We find that with one wearing session used for training (40 seconds of data), EarCapAuth achieves an identification accuracy of 39.63\% $\rm\pm$ 3.01\%. 
Increasing the number of wearing sessions during training to up to 11 significantly improves the identification accuracy up to 89.95\% $\rm\pm$ 5.17\%. 
To simulate participants taking out the device repeatedly to insert them for a short duration during enrollment, we tested using 3 seconds from 10 wearing session during training (approximately 30 seconds enrollment). 
We find that this achieves almost the same performance as using the full wearing session during training (85.93\% $\rm\pm$ 3.21\%).
After all, the positioning and, therefore, the capacitive electrode readings are consistent within wearing sessions, and as a result, using more data from one wearing sessions only slightly improves performance.

\subsubsection{Impact of Movement} To test wether the model is robust against motion-induced noise, we trained the identification model on the 12 wearing sessions of each user while they were at rest and used the data recorded while they moved as a test set (8 wearing sessions per user). The performance is only slightly worse, achieving an accuracy of 86.40\% which is even within the standard deviation of the performance achieved while the users were at rest. This suggests that EarCapAuth is relatively stable against motion artifacts caused by walking.

\section{Discussion}
\label{sec:discussion}
We have already evaluated and discussed performance with respect to the number of wearing sessions and the impact of motion on authentication and identification; in the following, we discuss more general aspects of EarCapAuth including potential attack vectors, how to deal with false rejections, and how EarCapAuth could be implemented into off-the-shelf earphones.

\subsection{Attack Vectors and Manipulation}
Although we see erroneous or opportunistic misuse (e.g., by family members) as the most common scenario, we discuss three different attack methods that could be used to gain unauthorized access via EarCapAuth.

\subsubsection{Ear Shape Replication}
An attacker could potentially attempt an ear shape replication attack by creating an accurate mold of a legitimate user’s ear canal. This might be achieved through a physical impression or by reconstructing the shape from images. However, in contrast to the attack surface of optical methods, the replica would also need to account for the unique characteristics of the skin, bones, and other anatomical features, all of which influence the capacitive values. These factors make such an attack far less practical for real-world implementation.

\subsubsection{Brute Force} One potential attack vector is brute force, where an attacker could attempt to emulate all combinations of capacitive readings by attaching capacitors to the system. Although the potential solution space is quite large, it can be reduced due to the limited variation in ear canal sizes across individuals. Nonetheless, this brute-force approach is deemed impractical given the high complexity and time required to explore all configurations. Time delays can serve as easy countermeasures to such attacks.

\subsubsection{Replay Attacks} Replay attacks represent another potential threat. Here, an attacker might record capacitive measurements using another manipulated eartip and attempt to replay the readings by attaching suitable capacitors to the electrodes. While this attack is the most promising, it is also the most difficult to disguise. An additional effective countermeasure could be to integrate EarCapAuth with other sensing modalities that ensure the device is worn by a user (e.g., detecting the pulse of the wearer through in-ear sounds \cite{butkow2023heart}).

\subsection{Comparison to Related Work}
A common principle proposed in the related work to perform authentication with earables is to measure ear canal sound reflections. 
In the audible range, this principle achieved an equal error rate of 14.9\%, which is worse than EarCapAuth \cite{derawi2017biometric}.
However, the performance of biometric principles, which use reflections of inaudible frequencies in the ear canal is better than EarCapAuth, achieving EER of as low as 1\% \cite{akkermans2005acoustic}.
Another authentication principle based on the variable leakage current of laptops \cite{ding2021leakage} measured by earables achieved 8\% false acceptance rate at 10\% false rejection rate, which is outperformed by EarCapAuth.

To the best of our knowledge, there have been no large-scale studies of the biometric identification and authentication performance of consumer devices. The NIST \cite{watson2009slapssegii} on professional large 3- and 2-inch fingerprint biometric authentication systems can therefore be considered a gold standard for such devices. Assuming a typical range of 95\%-99\% correct authentication (1\%-5\% false acceptance), 4 out of 5 (3-inch) or 3 out of 5 fingers are correctly authenticated. As the NIST study was much larger than our study, a direct comparison is difficult. 
In addition, Apple claims that the chances of misclassification with "FaceID is less than 1 in 1,000,000 with a single enrolled appearance", which suggests a much higher accuracy than our approach \cite{etherington2017face}.

\subsection{Wearable Implementation}
A critical challenge for the widespread adoption of EarCapAuth is the integration of capacitive electrodes into existing soft eartips, e.g., made from flexible silver PDMS \cite{fang2020flectile}. Encouragingly, recent patents from companies like Apple have proposed the embedding of EEG electrodes into the eartips of earphones. These electrodes could support capacitive sensing.

For a real-world deployment of EarCapAuth, a more sophisticated pre-trained model should only require data from the target user for enrollment. The model would then be fine-tuned directly on the earphones without the need for data from non-target users.
In addition, real systems should include auditory feedback from the earphones on the authentication decision. 
Especially if authentication fails, suitable auditory feedback could encourage the user to reposition the device slightly (similar to how fingerprint authentication encourages users to re-scan after failure). We think this is still useful because we deliberately keep the reading time short (0.33 s) so that positive or negative feedback is given immediately. Upon successful authentication, the previously rejected data could be used to further refine the model.

\section{Conclusion}
EarCapAuth presents a novel and robust approach for biometric authentication and identification using capacitive sensing embedded within earables. 
Users can quickly enroll in EarCapAuth by inserting them briefly into the ear canal multiple times.
Authentication and identification are performed on a short (0.33s) window.
By leveraging the unique anatomical properties of the ear canal, our evaluation shows that EarCapAuth achieves reliable performance with low error rates in authentication and identification across 20 participants, even under realistic motion-induced noise. 
We open-source the capacitive sensing eartip hardware of EarCapAuth under MIT license.
In combination with other biometric principles, EarCapAuth has the potential to offer secure access to sensitive information and enhanced user privacy in earable devices.


\bibliographystyle{ACM-Reference-Format}
\bibliography{bibliography}


\end{document}